\documentclass[iop]{emulateapj}

\usepackage{hyperref}
\usepackage{xcolor}
\usepackage{url}
\usepackage{amssymb}
\hypersetup{
    colorlinks,
    linkcolor={red!50!black},
    citecolor={blue!50!black},
    urlcolor={black} 
}
\usepackage{mathrsfs}
\usepackage{nicefrac}

\newcommand{\feh}{\ensuremath{[\mbox{Fe}/\mbox{H}]}}
\newcommand{\Kperror}{7}

\shorttitle{Thermal Spectrum of KELT-2Ab}
\shortauthors{Piskorz et al.}

\begin{document}

\title{Ground- and Space-Based Detection of the Thermal Emission Spectrum\\of the Transiting Hot Jupiter KELT-2Ab}

\author{Danielle Piskorz\altaffilmark{1}, Cam Buzard\altaffilmark{2}, Michael R. Line\altaffilmark{3}, Heather A. Knutson\altaffilmark{1}, Bj{\"o}rn Benneke\altaffilmark{4},\\ Nathan R. Crockett\altaffilmark{1}, Alexandra C. Lockwood\altaffilmark{5}, Geoffrey A. Blake\altaffilmark{1,2}, Travis S. Barman\altaffilmark{6},\\ Chad F. Bender\altaffilmark{7}, Drake Deming\altaffilmark{8}, John A. Johnson\altaffilmark{9}}

\altaffiltext{1}{Division of Geological and Planetary Sciences, California Institute of Technology, Pasadena, CA 91125}
\altaffiltext{2}{Division of Chemistry and Chemical Engineering, California Institute of Technology, Pasadena, CA 91125}
\altaffiltext{3}{School of Earth and Space Exploration, Arizona State University, Tempe, AZ 85287} 
\altaffiltext{4}{Institute for Research on Exoplanets,  Universit{\'e} de Montr{\'e}al, Montreal, Canada}
\altaffiltext{5}{Space Telescope Science Institute, Baltimore, MD 21218}
\altaffiltext{6}{Lunar and Planetary Laboratory, University of Arizona, Tucson, AZ 85721}
\altaffiltext{7}{Department of Astronomy and Steward Observatory, University of Arizona, Tucson, AZ 85721}
\altaffiltext{8}{Department of Astronomy, University of Maryland, College Park, MD 20742}
\altaffiltext{9}{Harvard-Smithsonian Center for Astrophysics; Institute for Theory and Computation, Cambridge, MA 02138}

\begin{abstract}
We describe the detection of water vapor in the atmosphere of the transiting hot Jupiter KELT-2Ab by treating the star-planet system as a spectroscopic binary with high-resolution, ground-based spectroscopy.  We resolve the signal of the planet's motion with deep combined flux observations of the star and the planet. In total, six epochs of Keck NIRSPEC $L$-band observations were obtained, and the full data set was subjected to a cross correlation analysis with a grid of self-consistent atmospheric models. We measure a radial projection of the Keplerian velocity, $K_P$, of 148 $\pm$ \Kperror\ km s$^{-1}$, consistent with transit measurements, and detect water vapor at 3.8$\sigma$. We combine NIRSPEC $L$-band data with \textit{Spitzer} IRAC secondary eclipse data to further probe the metallicity and carbon-to-oxygen ratio of KELT-2Ab's atmosphere. While the NIRSPEC analysis provides few extra constraints on the \textit{Spitzer} data, it does provide roughly the same constraints on metallicity and carbon-to-oxygen ratio. This bodes well for future investigations of the atmospheres of non-transiting hot Jupiters. 
\end{abstract}


\section{Introduction}
Thousands of extrasolar planets have been discovered by surveys using the transit, radial velocity (RV), direct imaging, and microlensing methods. Of these planets, the ones most ripe for direct follow-up observations are those discovered by the transit method. If we wish to measure the atmospheric compostion of an exoplanet, we are typically limited to space-based measurements of the hottest and largest transiting planets. Transit spectroscopy and eclipse spectroscopy have successfully measured the atmospheres of hot Jupiters (planets the size of Jupiters located within 0.05 AU of their stars) and some mini Neptunes and super Earths. These techniques reveal the presence of water vapor, CO$_2$, CH$_4$, and other species in exoplanetary atmospheres (e.g., \citealt{Madhu2012}). They also provide insight into the atmospheric temperature-pressure structure (e.g., \citealt{Knutson2008}) and into the presence and behavior of clouds or hazes (e.g., \citealt{Sing2016}). However, transit photometry is a broadband measurement and is thus incapable of resolving molecular bands, resulting in degeneracies in retrieved atmospheric molecular abundances.


In contrast, high signal-to-noise, high-resolution spectroscopy provides a distinctly molecular approach to the study of hot Jupiter atmospheres.  These methods capitalize on the relative Doppler shift of a star's spectrum and that of the hot Jupiter, essentially treating the star and the hot Jupiter as if they were members of a spectroscopic binary. This ground-based capability has been implemented in many studies and aims to measure the line-of-sight Keplerian velocity  $K_P$ of the hot Jupiter.  The technique has been applied at VLT/CRIRES (e.g. \citealt{Snellen2010}), Keck/NIRSPEC (e.g., \citealt{Lockwood2014}), ESO/HARPS (e.g., \citealt{Martins2015}), and CFHT/ESPaDOnS (e.g., \citealt{Esteves2017}) to study almost ten hot Jupiters. 

In the VLT/CRIRES program, systems are typically observed over a $\sim$half night when the change in the planet's line-of-sight velocity is the largest. This technique has provided high significance detections of various species in hot Jupiter atmospheres (e.g. \citealt{Birkby2013}), but it is fundamentally limited to rapidly moving exoplanets and will have an increasingly difficult time isolating distant planets whose single night radial velocity variations are small.

With the NIRSPEC instrument at the Keck Observatory, \cite{Lockwood2014} and others have used multiple hour-long snapshots of hot Jupiter spectra at different orbital phases and therefore different line-of-sight orbital velocities. NIRSPEC's cross-dispersed echelle format allows for the detection of many planet lines over many orders at high signal-to-noise. The combination of many epochs of NIRSPEC data provides a measurement of the line-of-sight Keplerian velocity $K_P$. This multi-epoch technique in combination with high contrast imaging will retain the ability to detect further separated planets, out to orbital periods of $\sim$weeks to months, and thus into the habitable zone regime. With $K_P$ in hand, one can endeavour to determine the presence of water vapor (e.g., detecton of deep water absorption lines on 51 Peg b by \citealt{Birkby2017}), carbon monoxide (e.g., measurement of a volume mixing ratio of 10$^{-5}$ for CO on $\tau$ Boo b by \citealt{Brogi2012}), winds (detection of 2 km/s high-altitude winds on HD 209458 b by \citealt{Snellen2010}), and planetary rotation rate (measurement of a 2-day rotational period for HD 189733 b by \citealt{Brogi2016}). In addition, when applied to non-transiting planets, a measurement of $K_P$ effectively breaks the mass-inclination degeneracy  that limits the study of RV planets \citep[e.g.][]{Brogi2012,Brogi2013,Brogi2014,Lockwood2014,Piskorz2016,Piskorz2017,Birkby2017}.

This method's reliance on the Doppler shifting of the planet's spectrum provides a pathway towards not only characterizing the atmospheres of non-transiting planets, but also constraining atmospheric models of transiting planets having additional broadband data. The combination of space-based, low-resolution spectra with ground-based, high-resolution spectra was carried out on the hot Jupiter HD 209458 b \citep{Brogi2017}. The data set suggested an oxygen-rich atmosphere (C/O $<$ 1 at 3.5$\sigma$) and sub-stellar metallicity (0.1-1.0 times stellar at 1$\sigma$), and provided tighter constraints on the molecular abundances of water vapor, carbon monoxide, and methane than either dataset alone could have.

Here, we apply the observational and cross-correlation techniques presented in \cite{Piskorz2016} to the transiting hot Jupiter KELT-2Ab. As compared to high dispersion observations that utilize nights with rapidly varying exoplanet radial velocities {\citep{Snellen2010,Brogi2012}}, this multi-epoch approach presents more challenging data analysis and cross-correlation requirements, but retains the ability to study both transiting and non-transiting systems and can be applied to exoplanets at substantially larger orbital distances. A key aspect of the present work is the development of a method for combining ground-based (Keck NIRSPEC) and space-based (Spitzer IRAC) transit observations to provide constraints on KELT-2Ab's atmospheric composition.  

KELT-2 (also commonly known as HD 42176) was targeted by the KELT (Kilodegree Extremely Little Telescope) North transit survey. Once the initial transit detection was made with five years' worth of data, follow-up radial velocity measurements were made with TRES (Tillinghast Reflector Echelle Spectrograph) and follow-up photometry was taken with four telescopes \citep{Beatty2012}. KELT-2 is a binary star system with a hot Jupiter orbiting KELT-2A. KELT-2A is a late F star having $T_{\mathrm{eff}}$ = 6148 K and $R_{*}$ = 1.836 $R_{\sun}$. KELT-2B is a K2 star and was shown to be bound by the photometry presented in \citealt{Beatty2012}. The two stars have a projected separation of 2.29'' or 295 $\pm$ 10 AU.  The binary system was more recently observed by \cite{Wollert2015}, and remains bound.  KELT-2Ab orbits KELT-2A. It is a hot Jupiter with a mass of 1.52 $M_J$, a mildly-inflated radius of 1.29 $R_J$, and orbital period of 4.11 days. The relevant properties of KELT-2A and KELT-2Ab are given in Table~\ref{systemproperties}. The atmospheric composition (e.g. C/O ratios) of gaseous planets such as hot Jupiters can be used as evidence in understanding their formation history \citep{Oberg2011}. KELT-2Ab is a particularly interesting target for atmospheric composition studies because it provides an example of hot Jupiter formation in a binary stellar environment. 

In Section~\ref{spitzobs}, we detail Spitzer observations of KELT-2Ab and reduction. Section~\ref{obsreduc} details the NIRSPEC observations of KELT-2Ab and reduction. Section~\ref{scchimera} describes the self-consistent grid of planetary atmospheric models used in Section \ref{models}'s cross-correlation analysis of the NIRSPEC data. We calculate a NIRSPEC-informed prior in Section~\ref{models} and use it to fit atmospheric models to Spitzer observations in Section~\ref{combo}. We discuss our measurements of the planet's atmosphere in Section \ref{discuss} and conclude in Section~\ref{conclude}.


\begin{deluxetable}{llc}[t]
\tablewidth{0pt}
\tabletypesize{\scriptsize}
\tablecaption{KELT-2A System Properties}
\tablehead{Property & Value & Ref.} 
\startdata
\sidehead{\textbf{KELT-2A}}
Mass, $M_{\star}$ & 1.314 $^{+0.063}_{-0.060}M_{\sun}$ & (1)  \\
Radius, $R_{\star}$ & 1.836 $^{+0.066}_{-0.046}R_{\sun}$ & (1) \\
Effective temperature, $T_{\mathrm{eff}}$ & 6148 $\pm$ 48 K & (1) \\
Metallicity, \feh &0.034 $\pm$ 0.78 & (1) \\
Surface gravity, $\log g$ & 4.030 $^{+0.015}_{-0.026}$ & (1) \\
Rotational velocity, $v \sin i$ & 9.0 $\pm$ 2.0 & (1) \\
Systemic velocity, $v_{sys}$ & -47.4 km/s & (2) \\
\textit{K} band magnitude, $K_{mag}$ & 7.35 $\pm$ 0.03 & (3) \\

\sidehead{\textbf{KELT-2A b}}
Velocity semi-amplitude, $K$ & 161.1 $^{+7.6}_{-8.0}$ m/s & (1) \\
Line-of-sight orbital velocity, $K_P$ &145 $^{+9}_{-8}$ km/s & (1) \\
(transit measurement) & & \\
Line-of-sight orbital velocity, $K_P$ &148 $\pm$ \Kperror\ km/s & (4) \\
(NIRSPEC measurement) & & \\
Mass, $M_p$ & 1.524 $\pm$0.088 $M_J$ & (1) \\
Radius, $R_p$ & 1.290$^{+0.064}_{-0.050}$ $R_J$ & (1) \\
Semi-major axis, $a$ & 0.05504 $\pm$ 0.00086 AU & (1) \\
Period, $P$ &4.1137913 $\pm$ 0.00001 days & (1) \\
Eccentricity, $e$ & 0 & (1) \\
Argument of periastron, $\omega$ & 90$^{\circ}$ & (1) \\
Time of periastron, $t_{peri}$ &   2455974.60338$^{+0.00080}_{-0.00083}$ JD &(1) \\
\enddata
\label{systemproperties}
\tablerefs{(1) \citealt{Beatty2012}, (2) \citealt{Gont2006}, (3) \citealt{Cutri2003}, (4) This work} 
\end{deluxetable}

\section{Spitzer Observations and Data Reduction}
\label{spitzobs}

\subsection{Spitzer Observations}
\begin{figure}[t]
\centering
\noindent\includegraphics[width=15pc]{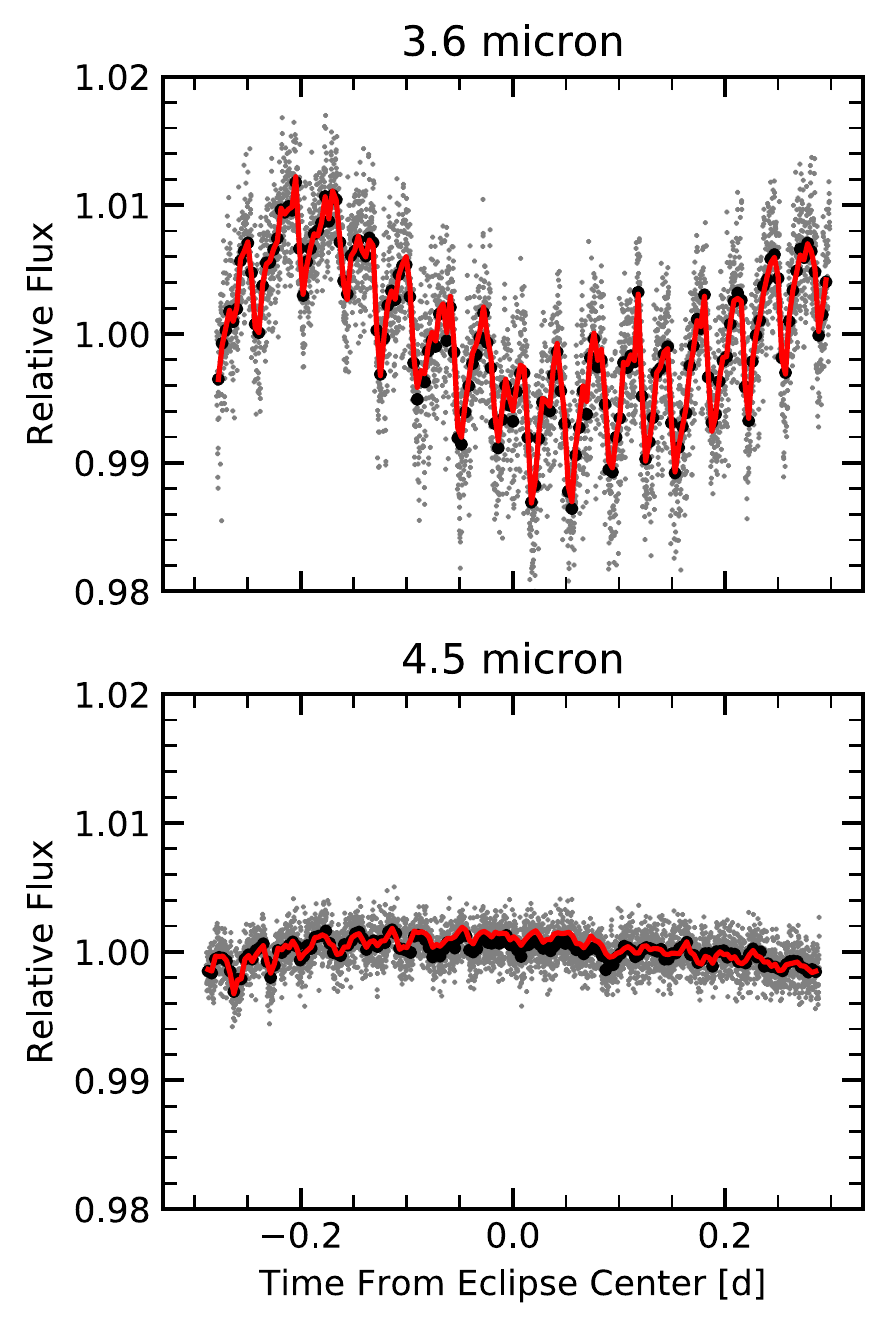}
\caption{Raw Spitzer photometry for 3.6 and 4.5 $\mu$m secondary eclipses plotted in 10-second (grey) and 5-minute (black) bins. The best-fit detector model for each observation is shown as a red line.}
\label{raw}
\end{figure}

We observed KELT-2A's secondary eclipse in the 3.6 and 4.5 $\mu$m bands for one session each with the Infrared Array Camera (IRAC; \citealt{Fazio2004}) on the Spitzer Space Telescope \citep{Werner2004} as a part of Program ID 10102 (Deming et al.). Spitzer observations and results are given in Table~\ref{spitztable}. We used the standard peak-up pointing mode for these observations, which places the star reliably in the center of a pixel after allowing for an initial 30 minute settling time at the new pointing position.  We observed our target in subarray mode with 0.4 s exposures in both bandpasses with a total duration of 14.4 hours (120,832 images) for each visit.  The raw photometry for each band pass is shown in Figure~\ref{raw} and the data with detector trends removed and best-fit light curves is shown in Figure~\ref{spitztransits}.

\begin{deluxetable*}{ccccccccc}
\tablewidth{0pt}
\tablecaption{Spitzer Observations and Measurements of KELT-2Ab}
\tablehead{$\lambda$ & Start Date& $t_{trim}^{\tablenotemark{a}}$ & $n_{bin}^{\tablenotemark{b}}$ & $r_{phot}^{\tablenotemark{c}}$& Background$^{\tablenotemark{d}}$& Eclipse Depth & Eclipse Time$^{\tablenotemark{e}}$ \\ 
($\mu$m) &  (UT) & (hr) & &  & (\%)& (ppm) & (BJD\_UTC)}
\startdata
3.6& 2014 Dec 17 &  0.5& 192& 2.5& 0.61 & 572$^{+45}_{-46}$ & 2457009.218$\pm$0.001\\
4.5& 2014 Dec 25 & 0.5 & 198 & 3.0& 0.38 & 616$^{+44}_{-45}$ & 2457017.448$\pm$0.001\\
\enddata
\label{spitztable}
\tablenotetext{a}{$t_{trim}$ is the amount of time trimmed from the start of each time series.}
\tablenotetext{b}{$n_{bin}$ is the bin size used in the photometric fits.}
\tablenotetext{c}{$r_{phot}$ is the radius of the photometric aperture in pixels.}
\tablenotetext{d}{Relative sky background contribution to the total flux in the selected aperture.}
\tablenotetext{e}{Eclipse times are consistent with a circular orbit.}
\end{deluxetable*}

\subsection{Spitzer Data Reduction}
We utilize the flat-fielded and dark-subtracted “Basic Calibrated Data” (BCD) images provided by the standard Spitzer pipeline for our analysis.  We first estimate the sky background by masking out a circle with a radius of 15 pixels centered on the position of the star, as well as the central several (13th-16th) columns and the central two (14th-15th) rows, which are contaminated by diffraction spikes from the star.  We also exclude the top (32nd) row of the array, which displays a systematically lower value than the rest of the image.  We then discard 3$\sigma$ outliers and make a histogram of the remaining pixel values, which are drawn from the corners of the 32$x$32 pixel array.  We fit this histogram with a Gaussian function to determine the sky background and subtract this background from each image.  

We determine the position of the star in each image using flux-weighted centroiding with a radius of 3.5 pixels, and calculate the flux in a circular aperture with radii of 2.0 - 3.0 pixels (in steps of 0.1 pixels) and 3.5 - 5.0 pixels (in steps of 0.5 pixels) to create our photometric time series.  We consider an alternative version of the photometry utilizing a time-varying aperture, where we scale the radius of the aperture proportionally to the square root of the noise pixel parameter, which is proportional to the full width half max (FWHM) of the stellar point spread function \citep{Knutson2012, Lewis2013}, but find that we obtain optimal results in both bandpasses using a fixed aperture.  In all cases, we calculate the noise pixel parameter using an aperture with a radius of 4.0 pixels. 

\begin{figure}[h]
\centering
\noindent\includegraphics[width=15pc]{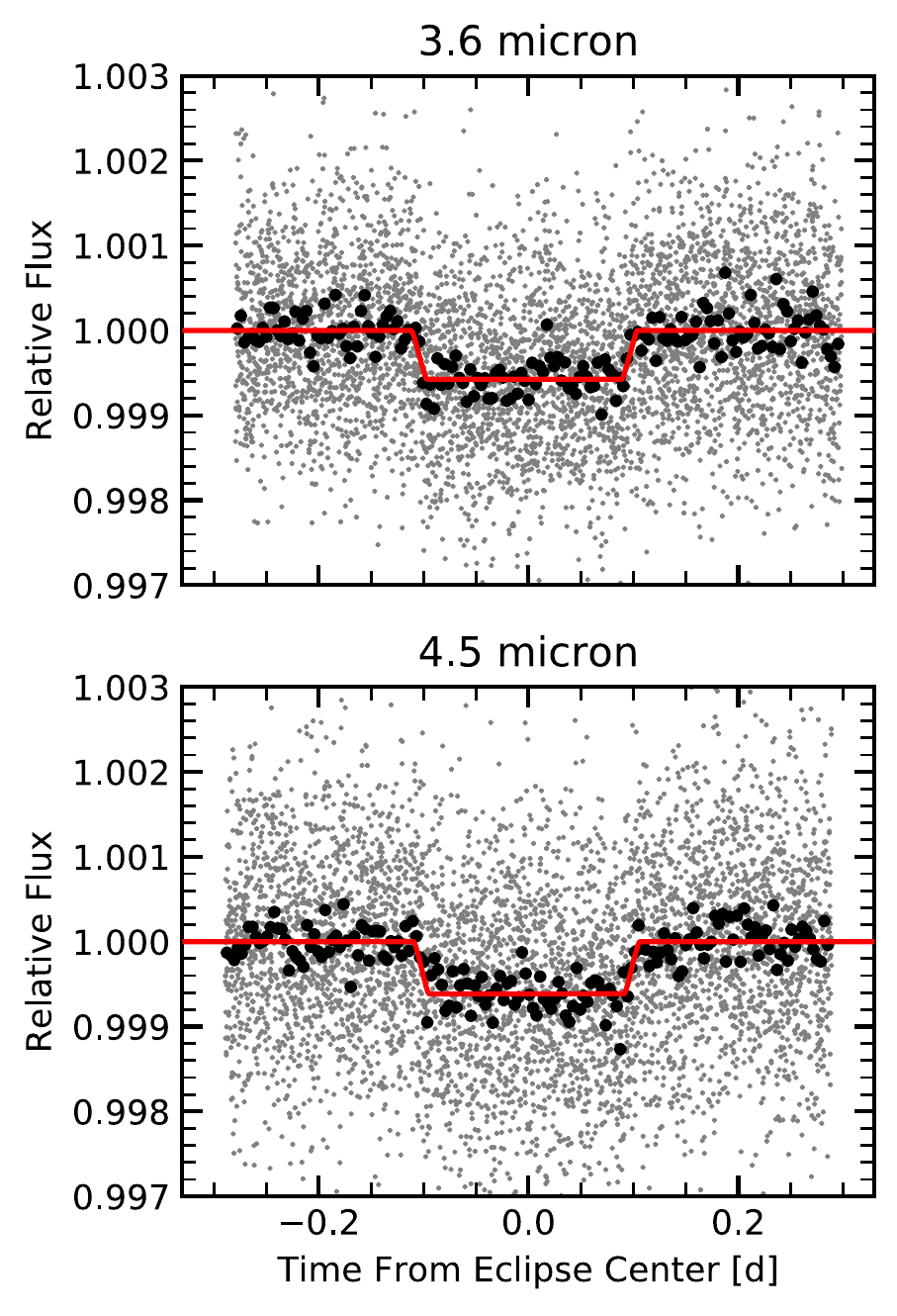}
\caption{Normalized secondary eclipse light curves after dividing out the best-fit detector noise model are shown in grey (10-second bins) and black (5-minute bins). The best-fit eclipse light curves are overplotted in red.}
\label{spitztransits}
\end{figure}

\begin{figure}[t]
\centering
\noindent\includegraphics[width=15pc]{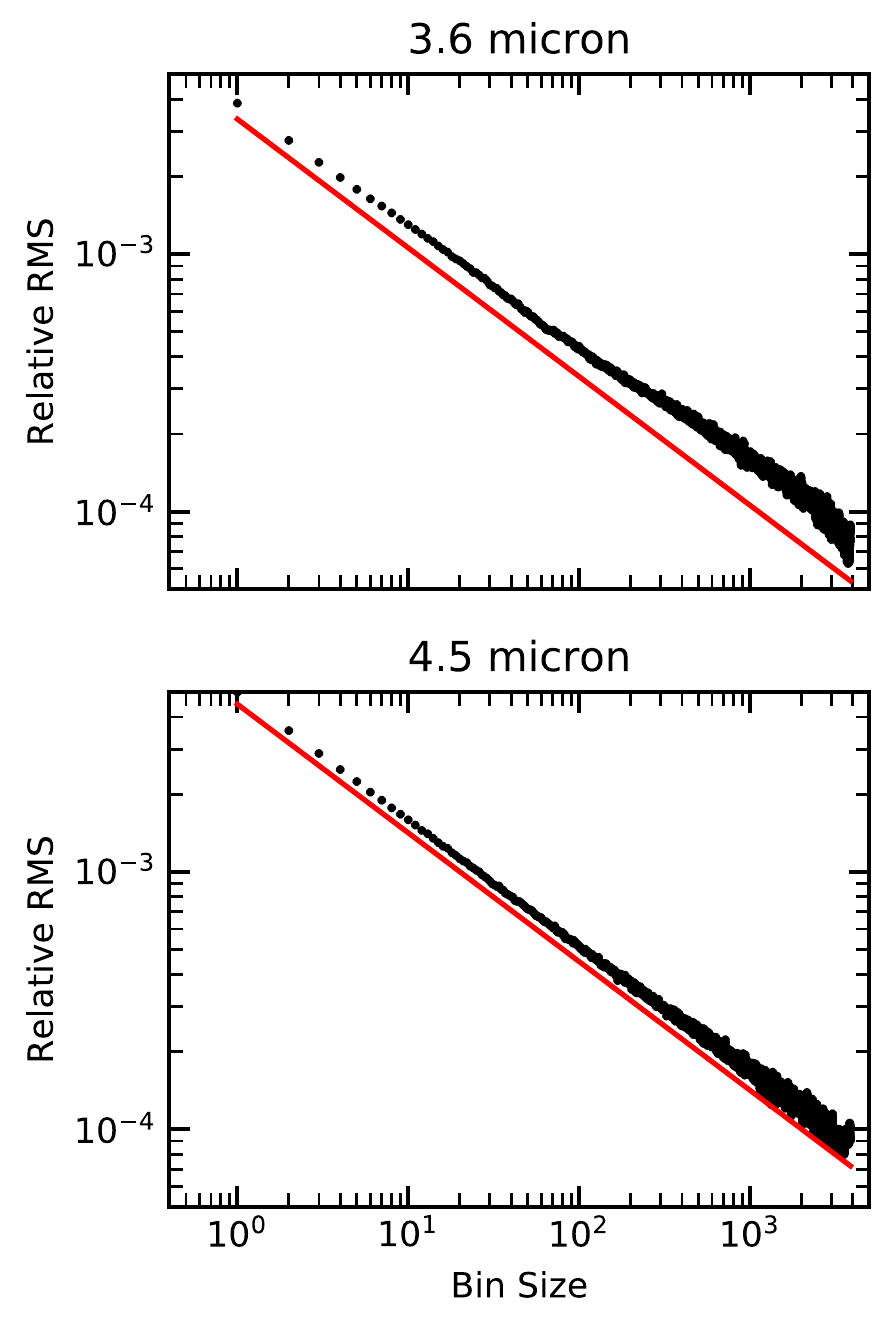}
\caption{Standard deviation of residuals for 3.6 and 4.5 $\mu$m Spitzer light curves as a function of bin size are shown in black with the predicted photon noise limit for each channel scaled by the square root of the number of points in each bin shown in red for comparison.}
\label{binsize}
\end{figure}

After extracting a photometric time series for each visit, we fit each time series with the pixel-level decorrelation (PLD) model described in \cite{Deming2015}, in which we utilize a postage stamp of nine pixels centered on the position of the star.  We also evaluate the need for a ramp using the the Bayesian Information Criterion (BIC), and find that it strongly favors the use of an exponential function in the 3.6 $\mu$m fit ($\Delta_{\textrm{BIC}}=-39$) but prefers a linear term in the 4.5 $\mu$m fit ($\Delta_{\textrm{BIC}}=+12$). Thus, we include both a linear and (for the 3.6 $\mu$m data only) an exponential function of time. We also assume that all points in our time series have the same measurement error, and allow this error to vary as a free parameter in our fits.  As discussed in \cite{Kammer2015} and \cite{Morley2016}, we optimize our choice of aperture, bin size, and trim duration individually for each visit by selecting the options which simultaneously minimize the RMS of the unbinned residuals as well as the time-correlated noise in the data.  We quantify this time-correlated noise component by calculating the RMS as a function of bin size in steps of 2$n$ points per bin (Figure~\ref{binsize}), and then take the least squares difference between the log of the predicted square root of $n$ scaling and the log of the actual RMS as a function of bin size.

In order to improve the convergence of our Markov-Chain Monte Carlo (MCMC) fits we elect to reduce the degrees of freedom in our model by using linear regression to determine the optimal set of nine PLD coefficients at each step in the MCMC chain.  Although this might cause us to under-estimate the uncertainties in our best-fit eclipse depth and time, we find that in practice the uncertainties in these parameters change by a negligible amount when we allow the PLD coefficients to vary as free parameters in our fits as compared to the linear regression approach.  It also has the added benefit of substantially reducing the convergence time for our MCMC chains, as the nine PLD coefficient values are strongly correlated with one another and it takes substantial time to fully explore this nine-dimensional space.

To fit the secondary eclipse light curves, we use the \texttt{batman} package \citep{Kreidberg2015}. Figure~\ref{spitztransits} shows the corrected Spitzer photometry and the best-fit secondary eclipse light curves for each channel. The 3.6 micron (channel 1) secondary eclipse depth is 572$^{+45}_{-46}$ ppm and the 4.5 micron (channel 2) secondary eclipse depth is  616$^{+44}_{-45}$ ppm. The eclipse times are consistent with a circular orbit. As seen in Figure~\ref{test}, the best fitting blackbody curve, at 1511 K, fits neither eclipse depth well ($\chi^2=8.4$), and so the Spitzer data are inconsistent with blackbody emission. 

These secondary eclipse measurements inform the contrast values used in the reduction and cross-correlation analysis of NIRSPEC $L$ band data discussed in Sections~\ref{obsreduc} and~\ref{models}. In Section~\ref{combo}, we will use these eclipse depths alone and in tandem with NIRSPEC $L$ band observations of KELT-2Ab to place constraints on the properties of the hot Jupiter's atmosphere.

\section{NIRSPEC Observations and Data Reduction}
\label{obsreduc}
\subsection{NIRSPEC Observations}
\label{nirspecobs}
\begin{figure}[t]
\centering
\noindent\includegraphics[width=20pc]{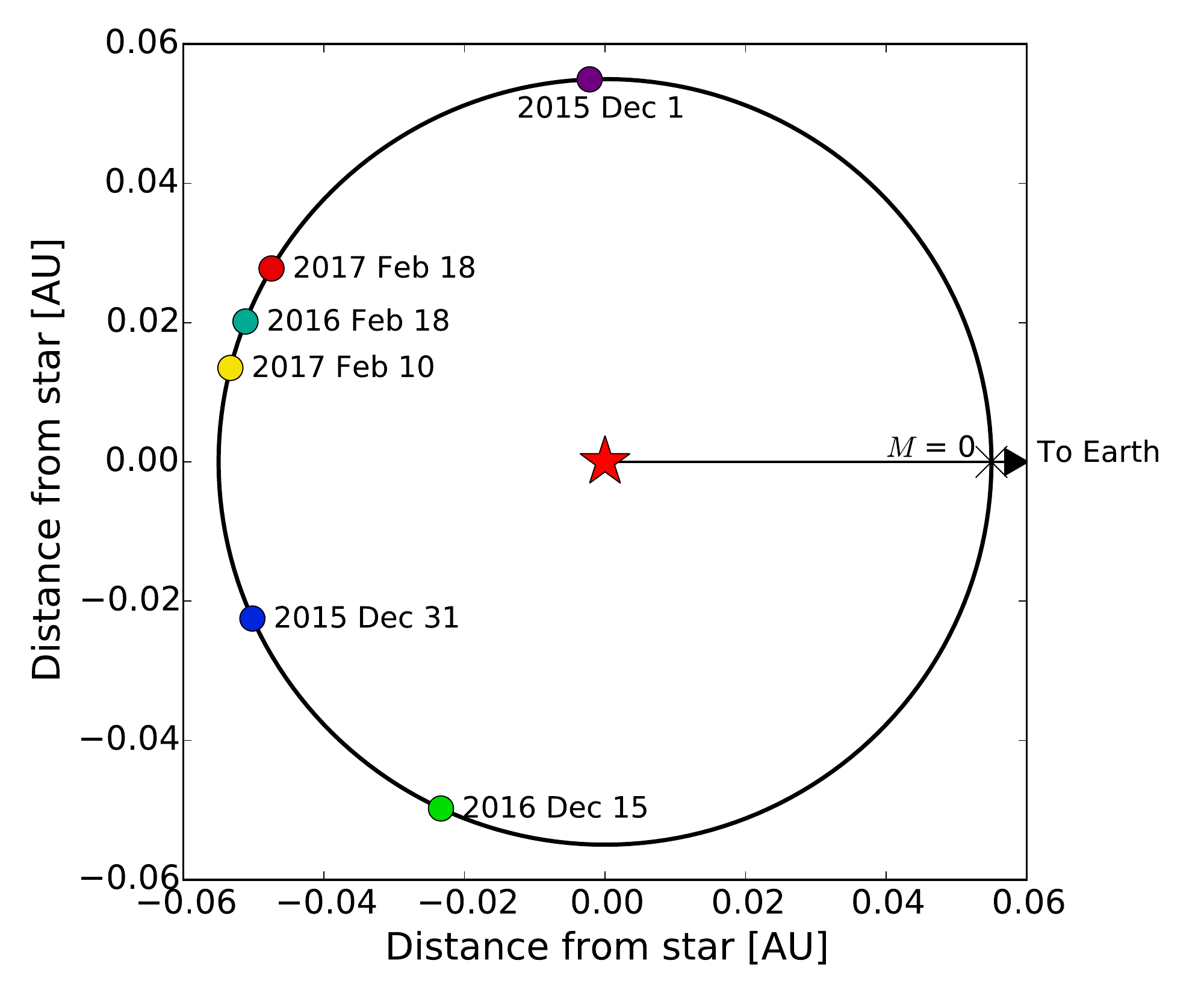}
\caption{Top-down schematic of the orbit of KELT-2Ab around its star according to the orbital parameters shown in Table~\ref{systemproperties}. Each point represents a single epoch of NIRSPEC observations of the system. The black arrow represents the line of sight to Earth.}
\label{schematic}
\end{figure}

\begin{figure}[t]
\centering
\noindent\includegraphics[width=20pc]{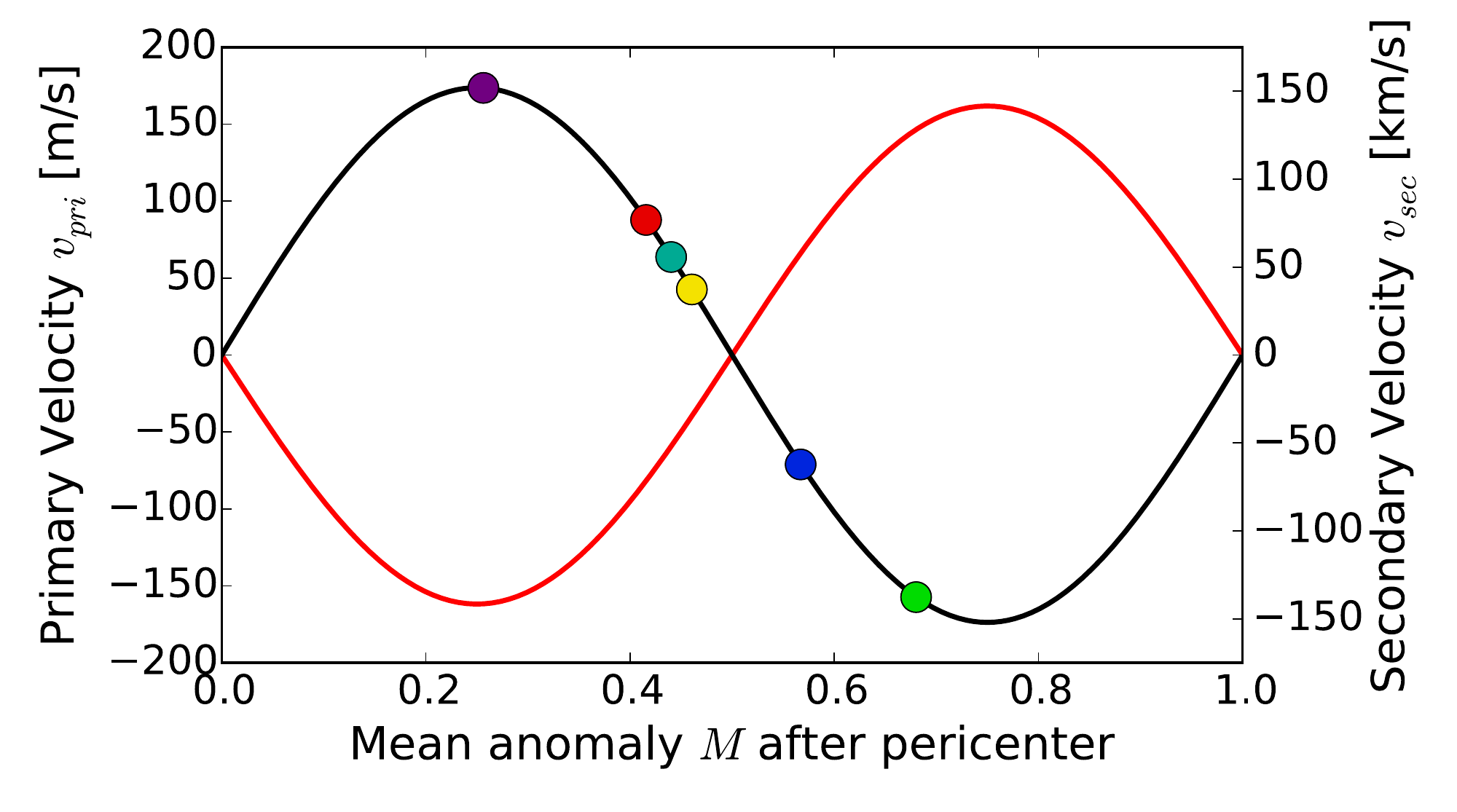}
\caption{Toy model showing the spectroscopic binary nature of the KELT-2A system. Based on values in Table~\ref{systemproperties}, the stellar RV curve is in red, and corresponds to the left ($v_{pri}$) y-axis (in m/s), and planetary RV curve is in black, corresponding to the right ($v_{sec}$) y-axis (in km/s). The colored points represent the NIRSPEC observations of this planet, correspond to the right y-axis, and are based on the observation phases and our expectations of the secondary velocities at those phases.}
\label{rvplot}
\end{figure}

We observe the KELT-2A system with NIRSPEC (Near InfraRed SPECtrometer; \citealt{McLean1998}) at Keck Observatory on six nights (2015 December 1, 2015 December 31, 2016 February 18, 2016 December 15, 2017 February 10, and 2017 February 18) in $L$ band. We use the 0.4''x24'' (3-pixel) slit setup, an ABBA nodding pattern for data acquisition (2-minute exposure per nod), and obtain resolutions (\textbf{$\nicefrac{\lambda}{\Delta\lambda}$}) of $\sim$ 25,000. We adjust the echelle and cross-disperser grating angles to provide wavelength coverage in each order of 3.4038-3.4565 / 3.2467-3.3069 / 3.1193-3.1698 / 2.995-3.044~$\mu$m. Table~\ref{observationtable} gives more details on these observations. The reported S/N's are the maximum achievable for each epoch of data. Functionally, however, potential error from residual telluric features, correlation with the stellar signal, etc., in addition to shot noise, put a ceiling on the S/N at which the planet can be detected. Figures~\ref{schematic} and~\ref{rvplot} provide the location and radial velocity of KELT-2Ab for each observation epoch. We observe the system when the planet has a high line-of-sight velocity and thermal emission from its hot dayside is visible. Our observations are short enough that the planet's signal does not smear across pixels for the entire co-added observation.

\begin{deluxetable*}{lccccc}
\tablewidth{0pt}
\tablecaption{NIRSPEC Observations of KELT-2A b}
\tablehead{Date & Julian Date$^{\tablenotemark{a}}$ & Mean anomaly $M^{\tablenotemark{a}}$ & Barycentric velocity $v_{bary}$ & Integration time & S/N$^{\tablenotemark{b}}_{\textit{L}}$ \\
 & (- 2,400,000 days) &  (2$\pi$ rad)  & (km/s) & (min) & }
\startdata
2015 December 1  & 57357.892 & 0.26 & 11.77 & 180& 1476\\
2015 December 31  & 57387.967 & 0.57 & -3.62 & 100 & 1125\\
2016 February 18  & 57436.810 & 0.44 & -24.74 & 80 & 1070\\
2016 December 15  & 57738.104 & 0.68 & 4.79 & 20 & 650\\
2017 February 10  & 57794.796 & 0.46 & -22.42 & 130 & 2103\\
2017 February 18  & 57802.867 & 0.42 & -24.93 & 140 & 1414\\
\enddata
\label{observationtable}
\tablenotetext{a}{Julian date and mean anomaly refer to the middle of the observing sequence.}
\tablenotetext{b}{S/N$_{\textit{L}}$ is calculated at 3.0 $\mu$m. Each S/N calculation is for a single channel (i.e., resolution element) for the whole observation.}
\end{deluxetable*}

\subsection{NIRSPEC Data Reduction}
\label{reduction}
We reduce our data and correct for telluric transmission with the Python pipeline from \cite{Piskorz2016}. In particular, the 2-D data are flat-fielded and dark subtracted according to \cite{Boogert2002}, while the 1-D spectra are extracted and wavelength calibrated with a fourth-order polynomial according to the position of the telluric lines. We fit and measure the instrument profile of our data following \cite{Valenti1995}.

\begin{figure}[t]
\centering
\noindent\includegraphics[width=21pc]{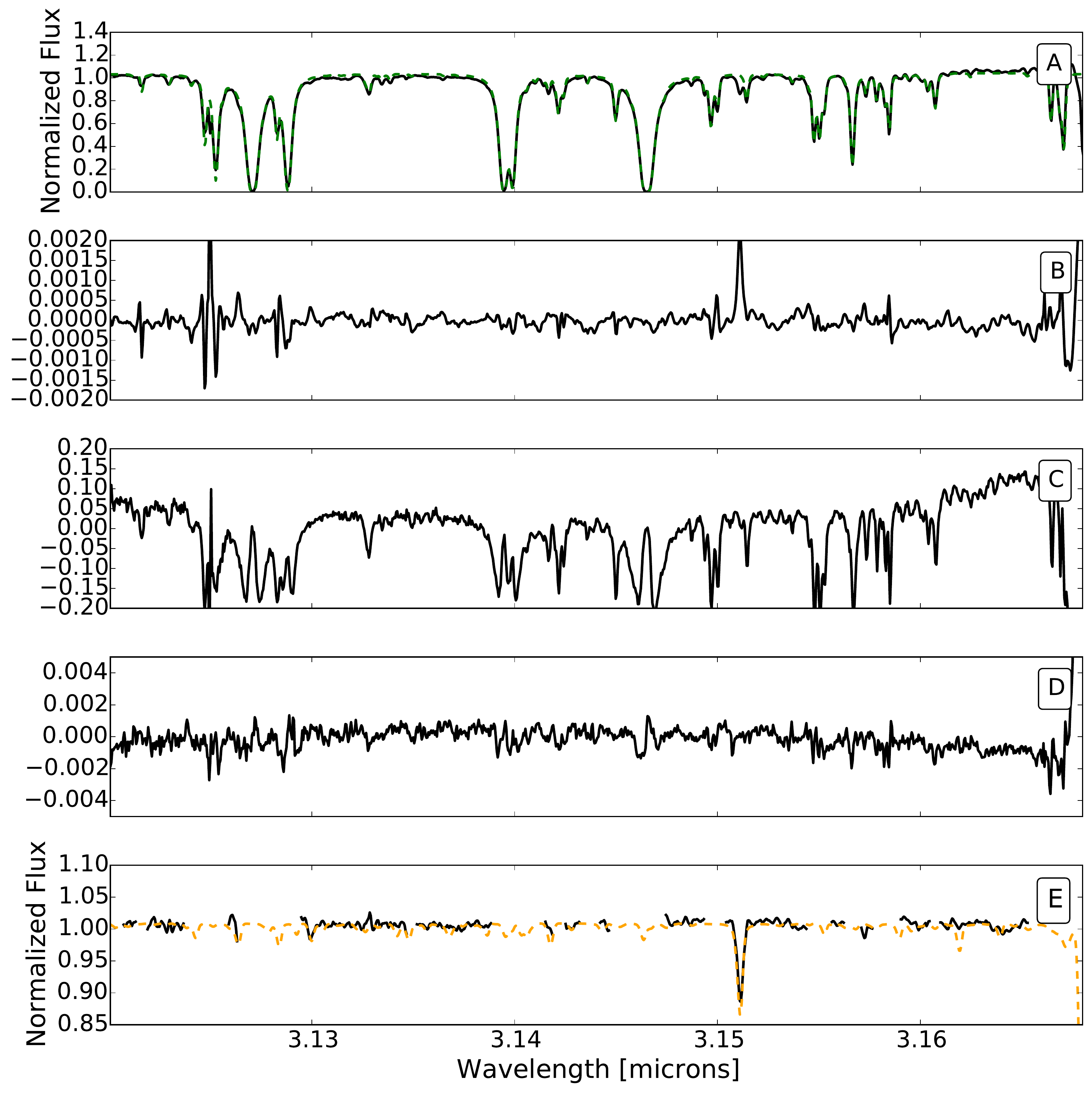}
\caption{Raw spectrum of KELT-2A, the first three principal components of the time-series of data, and a cleaned spectrum. (A): One order of data from KELT-2A taken on 2015 December 1. The best-fit telluric spectrum is over plotted as a dashed green line. (B-D): The first three principal components in arbitrary units. These components describe changes in air mass, molecular abundances in the Earth's atmosphere, and plate scale, respectively, over the course of the observation. (E): Same as (A), but without the first eight principal components and with saturated tellurics ($<0.75$) masked out. A fitted stellar spectrum is overplotted as a dashed orange line.}
\label{pcafigure}
\end{figure}

With a full set of 1-D spectra in hand, we use a model-guided principal component analysis (PCA) approach to remove tellurics and other time-varying signals from our data. For each epoch, we have a large time series of data, each of which can be rewritten as a linear combination of a set of basis vectors (i.e., principal components). The first few principal components capture the gross majority of the variance. This variance encapsulates all time-varying aspects of the data: changes in telluric abundances, changes in air mass, changes in the shape of the continuum, changes in the instrument response, etc. Removal of the strongest principal components from our data leaves behind the unchanging signal from the target star and the hot Jupiter. More information on our PCA approach is given in \cite{Piskorz2016} and a typical result of this analysis is shown in Figure~\ref{pcafigure}. 

After the first principal component is removed, removal of additional components makes little difference to the spectra, and the resulting correlation functions (described in Section \ref{models}) are roughly consistent with each other. We calculate the percent variance removed by each principal component, and find that, if the planet were moving over the course of a night (which we specifically select against observationally), we would still have to remove more than five to ten principal components to delete the signal from a typical hot Jupiter. As a sanity check, KELT-2Ab's expected photometric contrast $\alpha_{phot}$ suggested by the Spitzer data in Section~\ref{spitzobs} is roughly 600 ppm in the IRAC bandpasses, making it a typical hot Jupiter. In the following sections, we use a NIRSPEC dataset with the first eight principal components removed. 

This $L$ band data set will be interpreted with a two-dimensional cross-correlation technique (\cite{Zucker1994} and Section~\ref{models}). Such an analysis first requires a set of high-resolution planetary model spectra, which we describe in the next section.

\section{High-Resolution Atmospheric Models with S\lowercase{c}CHIMERA}
\label{scchimera}
We use a newly developed grid of cloud free and self-consistent thermochemical-radiative-convective equilibrium models (Self-consistent CHIMERA --- ScCHIMERA) to simultaneously interpret the Spitzer and NIRSPEC $L$ band data.  The CHIMERA framework was originally presented in \cite{Line2013}.

ScCHIMERA solves for radiative equilibrium using the \cite{Toon1989} two-stream source function technique for the planetary emission combined with a Newton-Raphson iteration scheme \citep{McKay1989}.  Opacities are treated within the ``resort-rebin" correlated-K (CK; \citealt{LO1990}) framework described in \cite{Molliere2015} and \cite{Amundsen2017}, and can handle any arbitrary combination of molecular abundances. The CK tables for H$_2$O, CH$_4$, CO, CO$_2$, NH$_3$, H$_2$S, HCN, C$_2$H$_2$, Na, K, TiO, VO, FeH, and H$_2$-H$_2$/He collision induced opacities are generated at an R=100 from 0.3 -200 $\mu$m for 20 Gauss-Quadrature g-ordinates from the line-by-line cross-section tables described in \cite{Freedman2008} and Table 1 of \cite{Freedman2014}. Extinction due to H$_2$ and He Rayleigh scattering is added in as a continuum absorber within the CK framework.  Convective adjustment while maintaining energy conservation is imposed where the radiative temperature gradient is steeper than the local adiabat.

\begin{figure}[h]
\centering
\noindent\includegraphics[width=21pc]{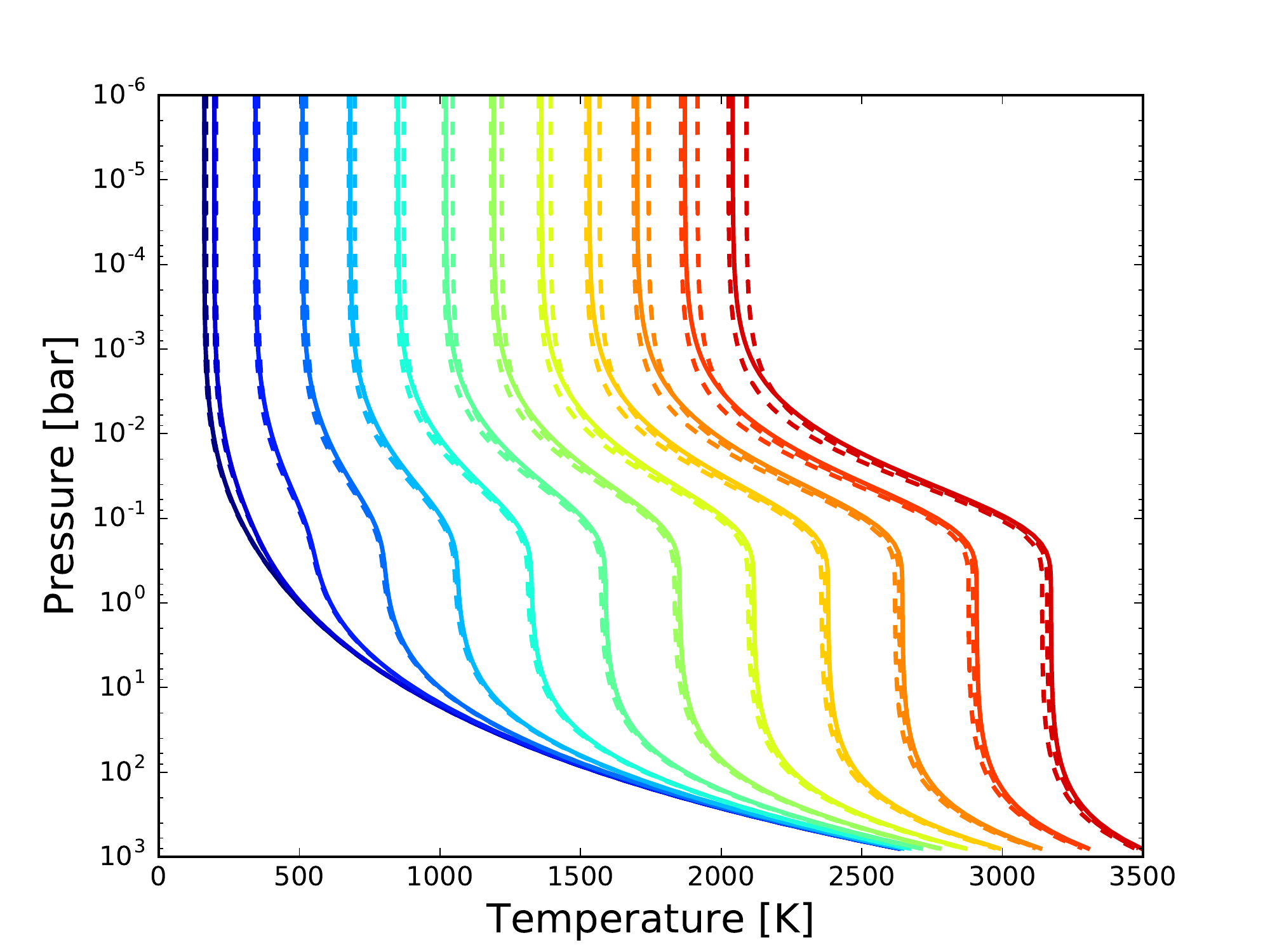}
\caption{Validation of irradiated temperature profiles derived from the ScCHIMERA numerical radiative-equilibrium solver (solid) against double-gray analytic solutions (dashed; from \citealt{Guillot2010}).  Analytic solutions for two different values of the radiative diffusivity are shown (e.g., \citealt{Parmentier2013}) and bracket the numerical solution which is exact in the limit of no-scattering \citep{Toon1989}.  The model set up here is for a gravity of 10 m/s$^2$, an internal temperature of 200 K, an infrared gray opacity of 0.3 m$^2$/kg,  a gray visible-to-infrared opacity ratio of 5$ \times 10^{-3}$, and a range of irradiation temperatures from 0 - 2300 K in steps of 200 K.  The numerical solution agrees with the analytic solutions to better than 3\% at all layers.}
\label{validation}
\end{figure}

\begin{figure}[h]
\centering
\noindent\includegraphics[width=21pc]{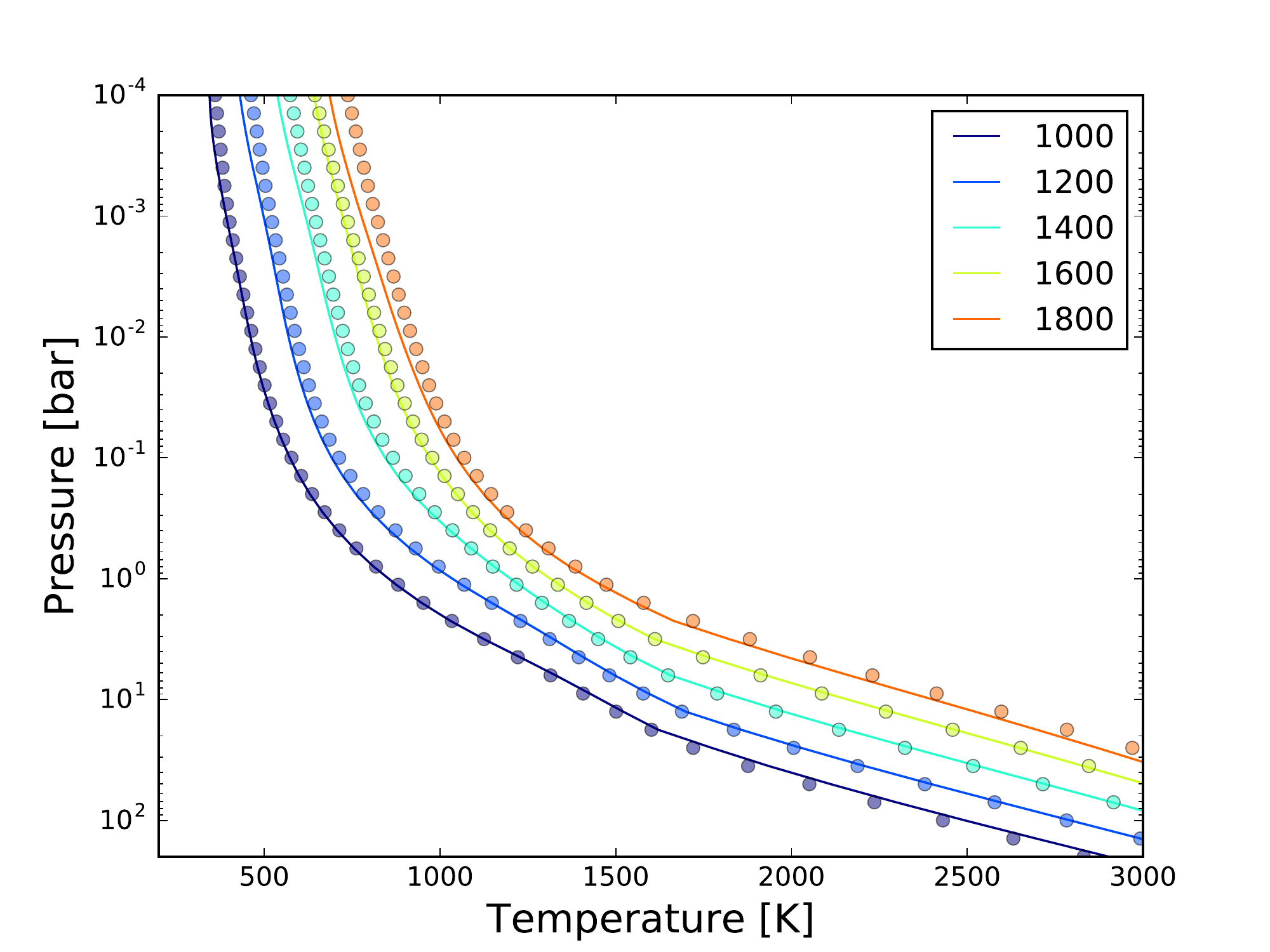}
\caption{Validation of a non-irradiated temperature profile using the ScCHIMERA (solid) correlated-K ``resort-rebin" opacity implementation against \cite{SM2008} grid models (dotted) for a $\log g$ of 5 (cgs) and effective temperatures from 1000-1800 K, for solar composition.  Differences are on the order of $\sim$5\%.  Differences may be attributed to the different treatment of the correlated-K opacities \citep{Amundsen2016}: ``pre-mixed" k-coefficients in \cite{SM2008} vs. ``resort-rebin" in this work.}
\label{opacity}
\end{figure}

In this cloud-free version we need not consider scattering in the visible stream. Here, we treat the wavelength-dependent incident stellar flux (from a PHOENIX stellar grid model; \citealt{Husser2013}) by including only the ``direct" beam and pure extinction over an average cosine zenith angle of $\nicefrac{1}{\sqrt{3}}$.  The incident stellar flux is scaled by a parameter, $f$, to account for day-night heat transport and an unknown albedo. Chemical equilibrium abundances, molecular weight, and atmospheric heat capacities are computed using the CEA2 routine \citep{Gordon1994} given the \cite{Lodders2003} abundances scaled via a metallicity and C/O ratio parameter (as in \citealt{Molliere2015}).  The radiative-convective numerical scheme is implemented in pure Python (with k-table $T$-$P$ interpolation in C) using the \texttt{anaconda numba} package for added acceleration.  Validation of the numerical implementation of the radiative-equilibrium solver against analytic solutions is shown in Figure~\ref{validation} and validation of the the opacity treatment against a brown dwarf grid model from \cite{SM2008} is shown in Figure~\ref{opacity}. Differences in all validation cases are on the order of 5\%. 

For the following cross-correlation analysis, we use a grid of ScCHIMERA planetary model spectra using the line-by-line version of the opacities given the converged T-P profile and thermochemical equilibrium molecular abundances. These spectra have resolution R = 500,000 and are calculated on the grid defined by metallicity $\log z$ = -1.0 - 2.0 in steps of 0.5, carbon-to-oxygen ratio C/O = 0.25 - 1.0 in steps of 0.25, and stellar flux scaling $f$ = 0.5 - 2.0 in steps of 0.25. Stellar flux scaling is a rough measure of energy redistribution. For $f$ $\gtrsim$ 1.5, the model atmosphere shows a temperature inversion. For the relevant regions of the grid, H$_2$O is the only significant $L$ band opacity source, and no model spectrum is consistent with a blackbody.

\section{NIRSPEC Data Analysis and Results}
\label{models}
%
%

\subsection{Two-Dimensional Cross Correlation}
\label{corr}
We measure the stellar and planetary velocity for each epoch of data with a two-dimensional cross correlation analysis (TODCOR), according to \cite{Zucker1994}, and with associated PHOENIX stellar and ScCHIMERA planet models. We use PHOENIX stellar model spectra based on the effective temperature, surface gravity, and metallicity of KELT-2A listed in Table~\ref{systemproperties} \citep{Husser2013}. In order to match the models to the observed spectra as closely as possible, all models are rotationally and instrumentally broadened before proceeding with the cross-correlation analysis.

\begin{figure}[t]
\centering
\noindent\includegraphics[width=21pc]{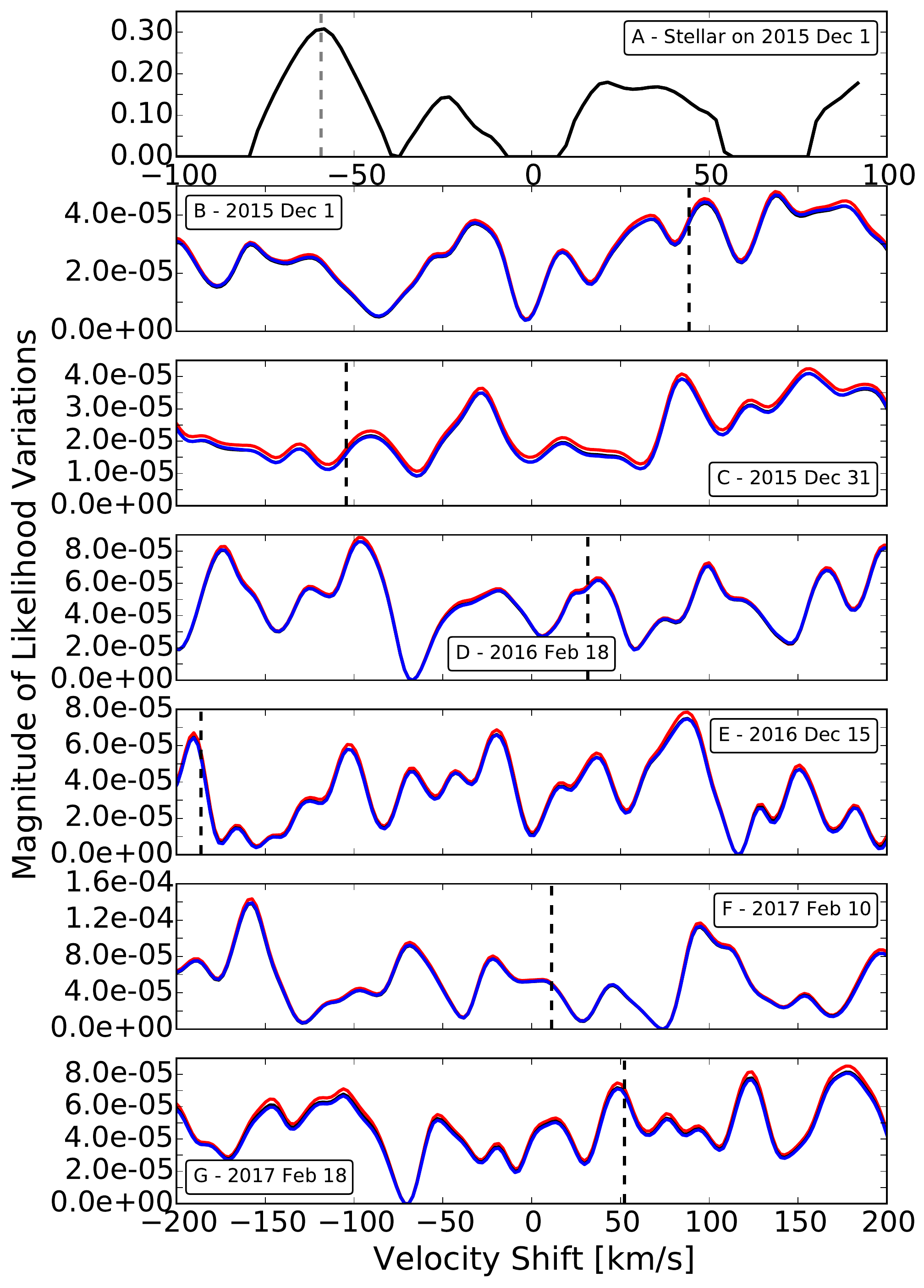}
\caption{Maximum likelihood functions for all epochs of $L$ band data. (A) Maximum likelihood function for the stellar velocity shift on 2015 December 1. (B-G) Maximum likelihood function for the planetary velocity shift.  The grey and black vertical lines represent the expected values of $v_{pri}$ and $v_{sec}$, respectively (based on the barycentric and systemic velocities and the line-of sight Keplerian velocity determined in Section \ref{orbitsoln}). The red, black, and blue curves represent the correlation with the NIRSPEC-only best-fit, Spitzer-only best-fit, and NIRSPEC+Spitzer best-fit planet models, respectively. Only by combining these likelihood curves do we detect the planet (see Figure~\ref{maxlike}).} 
\label{sxcorr}
\end{figure}

For each ScChimera model, TODCOR produces a matrix of correlation values for various stellar and planetary velocity shifts. We combine the correlation functions for the orders of a single epoch and calculate a nightly maximum likelihood curve for the star's and planet's velocities according to the relationship \citet{Lockwood2014}, who showed:
\begin{equation}
\log \mathcal{L} = \textrm{const} + CCF.
\end{equation}
An example of the resulting maximum likelihood curves for three sets of cross-correlations (each with a best-fitting ScCHIMERA model) is shown Figure~\ref{sxcorr}. With the PHOENIX model, we detect the star's velocity at a combination of the systemic velocity and the barycentric velocity, as expected (see Panel A of Figure~\ref{sxcorr}). This technique is not sensitive to the reflex motion of the star, which is below the velocity precision of NIRSPEC.

\subsection{Planet Mass and Orbital Solution}
\label{orbitsoln}
\begin{figure}[t]
\centering\noindent\includegraphics[width=20pc]{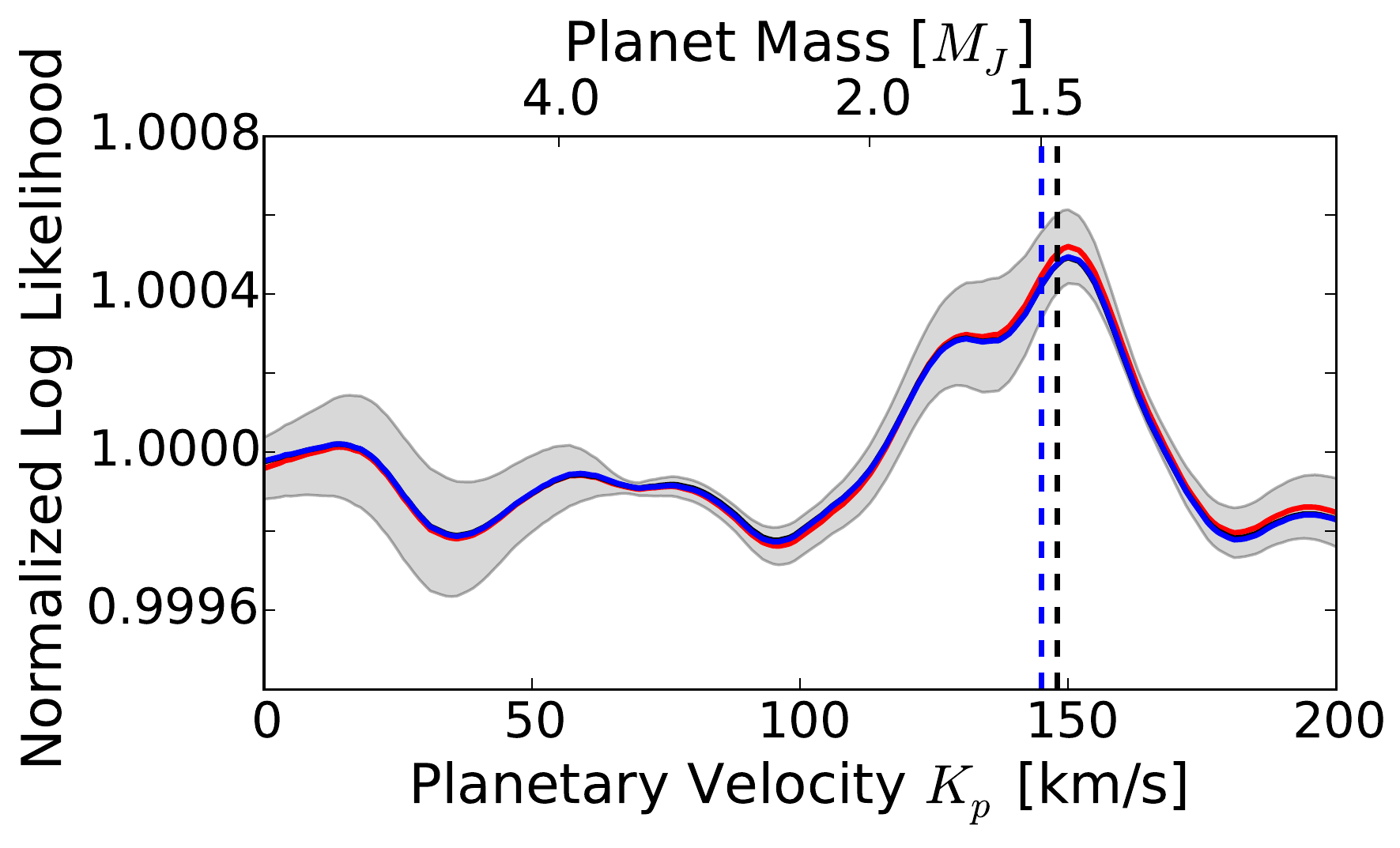}
\caption{Normalized log likelihood as a function of Keplerian orbital velocity $K_P$. Likelihood curves resulting from correlating NIRSPEC data with the NIRSPEC-only best-fit, Spitzer-only best-fit, and NIRSPEC+Spitzer best-fit planet models for KELT-2Ab in red, black, and blue, respectively. The grey shading represents the jackknifed error bars and the vertical dashed black line represents the detection of the planet's velocity at 148 $\pm$ \Kperror\ km/s. The vertical blue dashed line represents the measurements made by \cite{Beatty2012}.}
\label{maxlike}
\end{figure}

However, for a single epoch, we are unable to reliably identify the planet's velocity based on its nightly maximum likelihood curve (see Panels B-G of Figure~\ref{sxcorr}). To retrieve an estimate of the line-of-sight Keplerian velocity, we must combine the nightly maximum likelihood curves into a single, multi-epoch likelihood curve \citep{Lockwood2014}. Our equation for orbital velocity assumes a circular orbit, as is likely the case for KELT-2Ab \citep{Beatty2012}: 
\begin{equation}
\label{vf}
v_{sec}(M) = K_{p}\sin(2\pi M) + \gamma
\end{equation}
Here, $v_{sec}$ is the planet's velocity shift, $M$ is the mean anomaly of the observation epoch ($M$ = 0 at transit), and $\gamma$ is a combination of the systemic and barycentric velocities. We test a range of $K_P$ values from -250 to 250 km/s in order to create a final likelihood curve for each ScCHIMERA model as shown in Figure~\ref{maxlike}. 

For all ScCHIMERA models, we are able to detect the planet's velocity at 150 km/s. The line-of-sight Keplerian velocity  of KELT-2Ab is 145$^{+9}_{-8}$ km/s based on the transit method \citep{Beatty2012}. Our NIRSPEC measurement of the planet's velocity lies comfortably within this range.

\begin{figure}[t]
\centering\noindent\includegraphics[width=20pc]{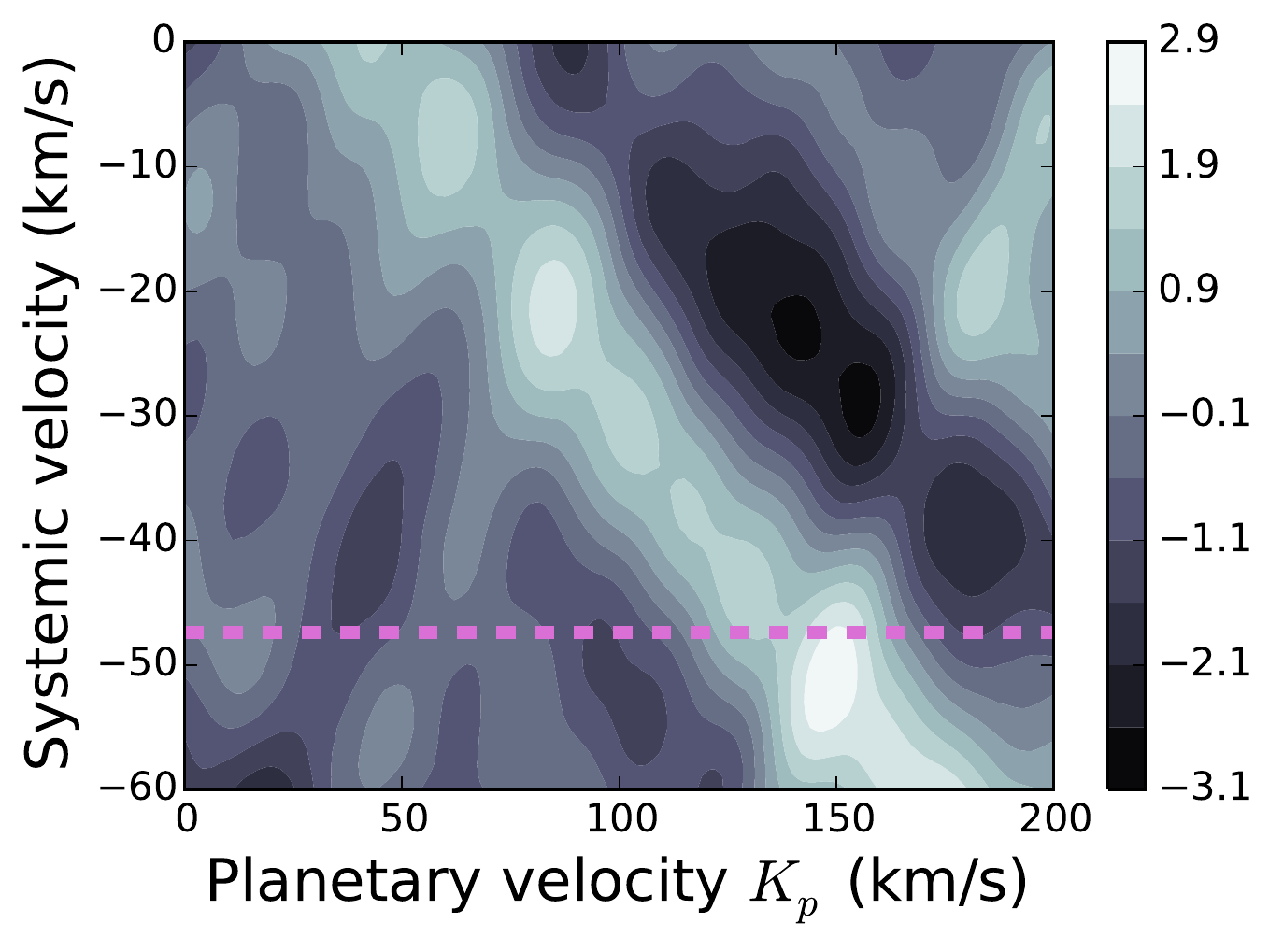}
\caption{ Normalized log likelihood as a function of Keplerian orbital velocity $K_P$ and systemic velocity $v_{sys}$. The contours indicate half $\sigma$ significance. The dashed magenta line shows the systemic velocity of -47.4 km/s as measured by \citealt{Gont2006}. }
\label{gammakp}
\end{figure}

In order to understand the significance of our planetary detection at the well-known systemic velocity, we consider the likelihood curve for $K_p$ at a range of different systemic velocities of the star (see Figure~\ref{gammakp}). The contours on this surface are given in terms of $\sigma$, where $\sigma$ is an approximation for the noise level given as the standard deviation of the full surface. Because the diagonal structure in the surface is non-random---it is caused by the degeneracy between $K_p$ and $v_{sys}$---this $\sigma$ statistic is not the best way of determining the significance of our detection and instead we use a technique that determines significance only from the true planetary peak, as described below.   

The NIRSPEC-only best fit likelihood curve shown in Figure~\ref{maxlike} is a cross section of the 2D surface along the known systemic velocity, which is represented by the dashed magenta line. The diagonal structure in this surface, as well as the shoulder to the main peak at $\sim$125 km/s, is caused by the degeneracy between $K_p$ and $v_{sys}$. The true maximum along the surface lies at a $v_{sys}$ of -51 km/s, but because this small offset is within the velocity resolution of NIRSPEC ($\sim$4 km/s), it is unlikely to be caused by any true physical process. Thus, we consider the significance of the peak in $K_p$ only along the cross section at the true systemic velocity of -47.4 km/s, as measured by \citealt{Gont2006}.

We calculate error bars on our measurement of $K_P$ with jackknife sampling. We remove one night's worth of NIRSPEC data, recalculate the likelihood curve, and repeat, resulting in six likelihood curves. The error bars for each value of $K_P$ are directly related to the standard deviation of the six likelihood curves as a function of $K_P$. These error bars are shown as grey shading in Figure~\ref{maxlike}.

With these error bars, we fit a Gaussian to the peak at 150 km/s, resulting in a $K_P$ measurement of 148 $\pm$ \Kperror\ km/s. This corresponds to a mass of 1.5 $\pm$ 0.1 $M_J$ and an orbital inclination of 79${^{+11}_{-9}}^{\circ}$. Based on this value of $K_P$, we mark the expected $v_{sec}$ for each observational epoch in Panels B-G of Figure~\ref{sxcorr}.

We determine the significance of the detection by comparing the likelihood of a Gaussian fit (indicating a planetary signal) and a linear fit (indicating no planetary signal) to the likelihood peak at 148 km/s. We calculate the Bayes factor $B$ as the ratio of likelihoods for the two fits. If 2ln$B$ is greater than ten, then the Gaussian fit is strongly preferred.

Using the jackknifed error bars, we find that 2ln$B$ is 11.6, suggesting that the planet detection is made at 3.8$\sigma$ \citep{Kass1995,Gordon2007}. We choose to determine the significance of our $K_p$ measurement this way because it only considers the real peak, and is not biased by the non-random structure at other values of $K_p$ (including the shoulder at $\sim$125 km/s which is due to the $v_{sys}$-$K_p$ degeneracy). For sufficiently deep integrations, this significance is determined by structure in the cross-correlation space and not by the aggregate shot noise.



\subsection{NIRSPEC Constraints on KELT-2Ab's Atmosphere}
\label{nirspecconstraints}
At $L$ band wavelengths, the planet model is dominated by water vapor and the source of the correlation signal presented here is water. Therefore, our NIRSPEC $L$ band data allows us to report the presence of water vapor in the atmosphere of KELT-2Ab at 3.8$\sigma$.

\begin{figure}[t]
\centering\noindent\includegraphics[width=20pc]{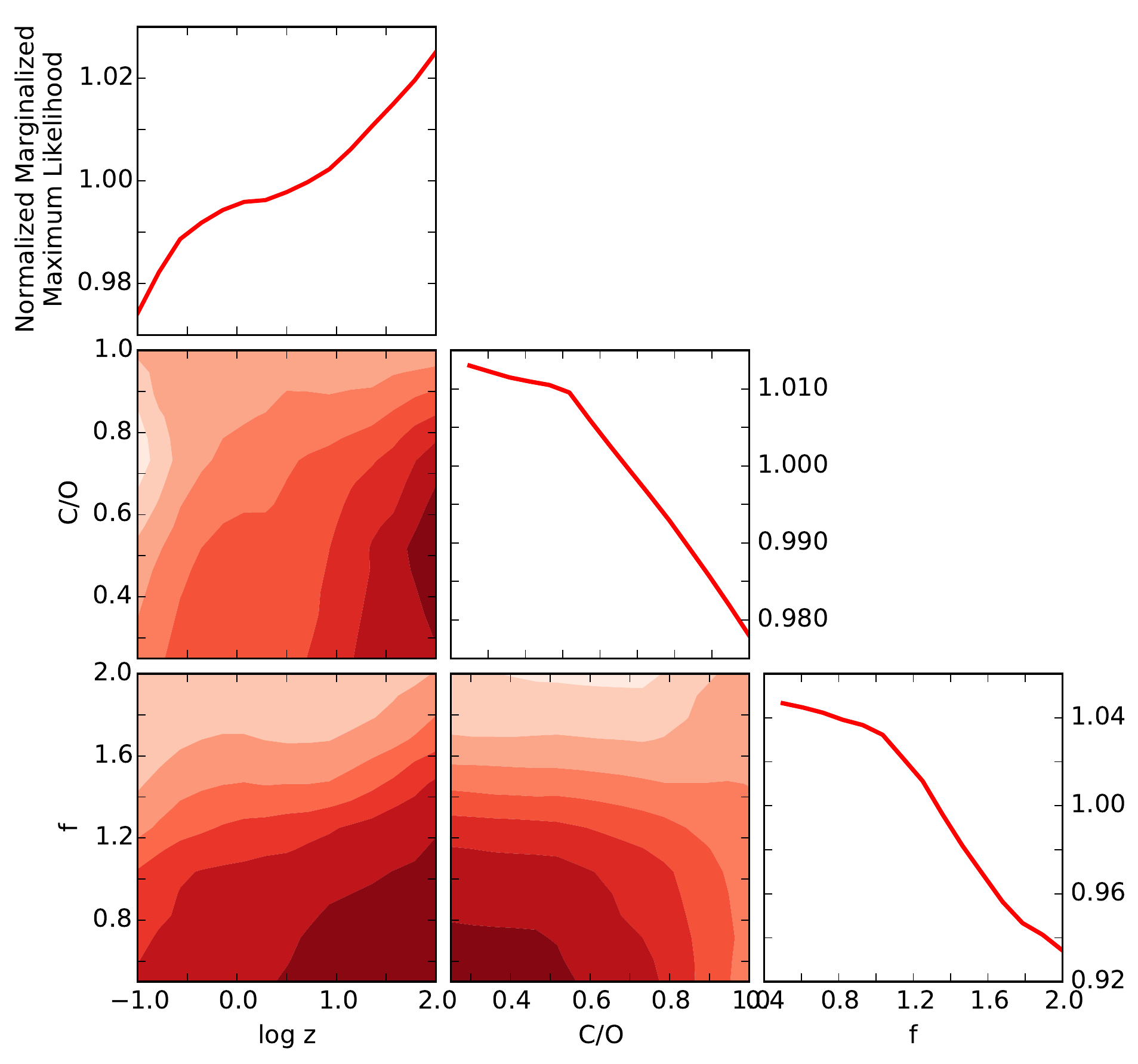}
\caption{NIRSPEC-only atmospheric fits results. The marginalized grid of ScCHIMERA models cross-correlated with NIRSPEC data is shown with regions of darker red indicating a higher likelihood. The line plots are the marginalized, normalized maximum likelihood values for each parameter. The grid of likelihoods is combined with the Spitzer MCMC likelihoods in Section~\ref{combo}.}
\label{nirspecprior}
\end{figure}

For each ScCHIMERA grid point, we record the normalized maximum value of the likelihood curve at $K_P$ = 150 km/s. We chose to record the likelihood values at 150 km/s because, for the best fitting model, while the Gaussian fit gives a central location of 148 km/s, the actual peak maximum is at 150 km/s. We also tested the likelihood values at 145 km/s, which is the $K_P$ measured from transit studies \citep{Beatty2012} and found that they gave the same result. To get an idea of the underlying structure of our calculated maximum likelihood grid, we marginalize the grid along each axis (Figure ~\ref{nirspecprior}).  The NIRSPEC data alone has preference for high metallicity, C/O $<$ 0.75, and low redistribution values. Specifically, the ScCHIMERA model that best fits the NIRSPEC data has $\log z =$ 1.5, C/O = 0.5, and $f$ = 1.0 (see Figure~\ref{test}). In Section~\ref{combo}, we combine the grid of NIRSPEC maximum likelihood values with the maximum likelihood values from the broadband Spitzer analysis of KELT-2Ab to find the joint best fitting models. 


\section{Joint Spitzer and NIRSPEC Constraints on KELT-2Ab's Atmosphere}
\label{combo}
Next we turn to the Spitzer secondary eclipse data introduced in Section~\ref{spitzobs} to further investigate KELT-2Ab's atmosphere. Specifically,
we use the ScCHIMERA model grid for KELT-2Ab discussed in Section \ref{models} at a resolution of R=100 to fit the Spitzer transit depths via the Markov-Chain Monte Carlo technique implemented in \texttt{emcee} \citep{FM2013}. The likelihood function is 
\begin{equation}
\mathscr{L} = \exp{\bigg(\frac{-(t_{obs}-t_{mod})^2}{2\sigma^2}\bigg)}
\end{equation}
where $t_{obs}$ is the observed transit depths, $t_{mod}$ is the transit depths of the model integrated over the Spitzer filters, and $\sigma$ is the error on the observed transit depth. Confined to the model grid, we initialize 50 chains, perform a burn-in of 2000 steps, and run each chain for an additional 10,000 steps. 

\begin{figure}[t]
\centering\noindent\includegraphics[width=20pc]{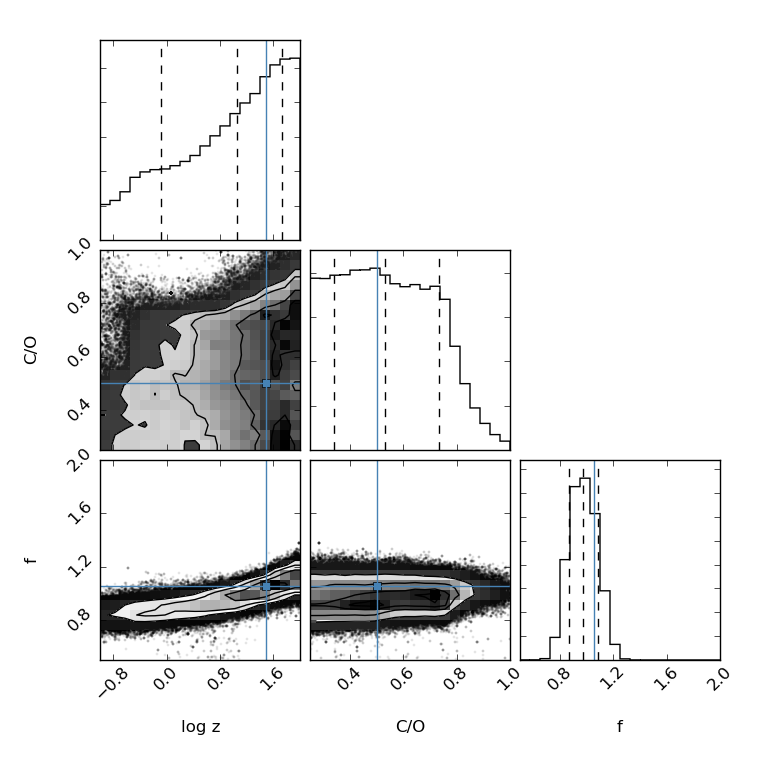} 
\caption{Spitzer-only atmospheric fit results. The results of an MCMC analysis of the Spitzer IRAC data points when fit with the ScCHIMERA models. Points are the best-fit models and the dashed lines are the 16, 50, and 84\% confidence intervals for each analysis.}
\label{cornerplot}
\end{figure}

\begin{figure}[t]
\centering\noindent\includegraphics[width=20pc]{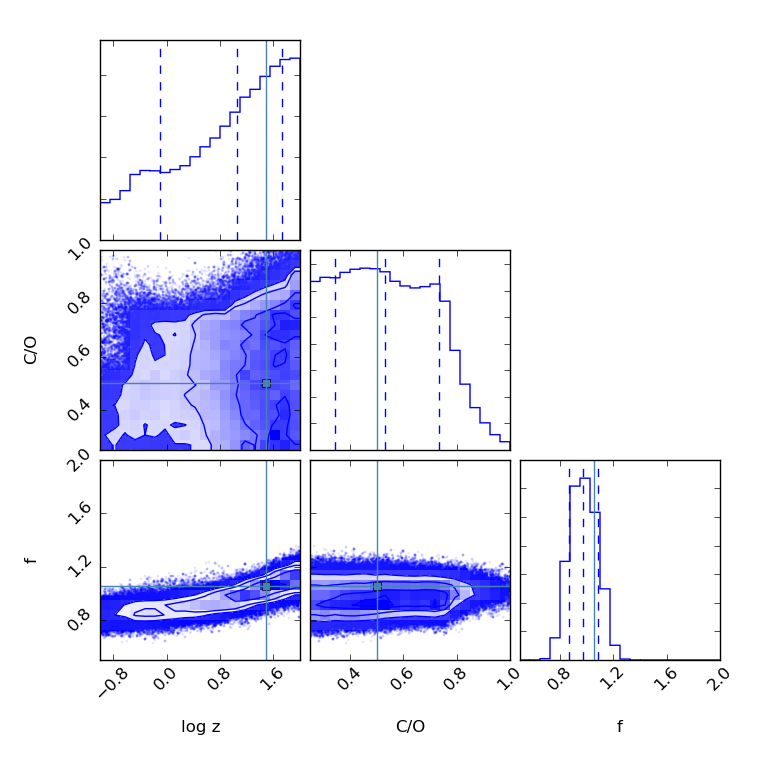} 
\caption{Combined Spitzer-NIRSPEC atmospheric fit results. The joint results of an MCMC analysis of the Spitzer IRAC (Figure~\ref{cornerplot}) and NIRSPEC data (which is essentially the likelihood grid illustrated in Figure~\ref{nirspecprior}) when fit with the ScCHIMERA models. Points are the best-fit models and the dashed lines are the 16, 50, and 84\% confidence intervals for each analysis.}
\label{jointcornerplot}
\end{figure}

We present this Spitzer MCMC analysis twice, each time with a uniform prior: once on its own (Figure~\ref{cornerplot}) and once combined with the likelihood surface calculated from the NIRSPEC cross-correlation analysis in Section~\ref{nirspecconstraints}, resulting in the corner plot shown in Figure~\ref{jointcornerplot} \citep{FM2016}. As the Spitzer and NIRSPEC measurements are independent of one another, the likelihoods can be multiplied. A Spitzer-prior on NIRSPEC, NIRSPEC-prior on Spitzer, and combined analysis are therefore all analogous. The best fit values and confidence intervals for both versions of the data fits are given in Table~\ref{bestfits}.

\begin{deluxetable}{lccccc}
\tablewidth{0pt}
\tablecaption{Best-Fit Values and Confidence Intervals for KELT-2A b Atmospheric Measurements}
\tablehead{Data Set & Parameter & 16\% CI & 50\% CI & 84\% CI & Best-fit}
\startdata
NIRSPEC,  & $\log z$ & -0.52 & 0.49& 1.51& 1.5\\
alone & $C/O$ & 0.37 &0.62 &0.88 & 0.5\\
  & $f$ & 0.74 & 1.26& 1.77& 1.0\\
\hline
Spitzer,  & $\log z$ & -0.10 & 1.05& 1.73&1.536 \\
alone & $C/O$ & 0.34 &0.53 &0.74 &0.403 \\
  & $f$ & 0.87 & 0.97& 1.08& 1.060\\
\hline
Joint Spitzer & $\log z$ & -0.11 &1.06 &1.73 &1.538 \\
and NIRSPEC & $C/O$ & 0.34 & 0.53& 0.73& 0.501\\
  & $f$ &  0.87& 0.97&1.08 &1.060 \\
\label{bestfits}
\end{deluxetable}

\section{Discussion}
\label{discuss}
\begin{figure*}[t]
\centering\noindent\includegraphics[width=40pc]{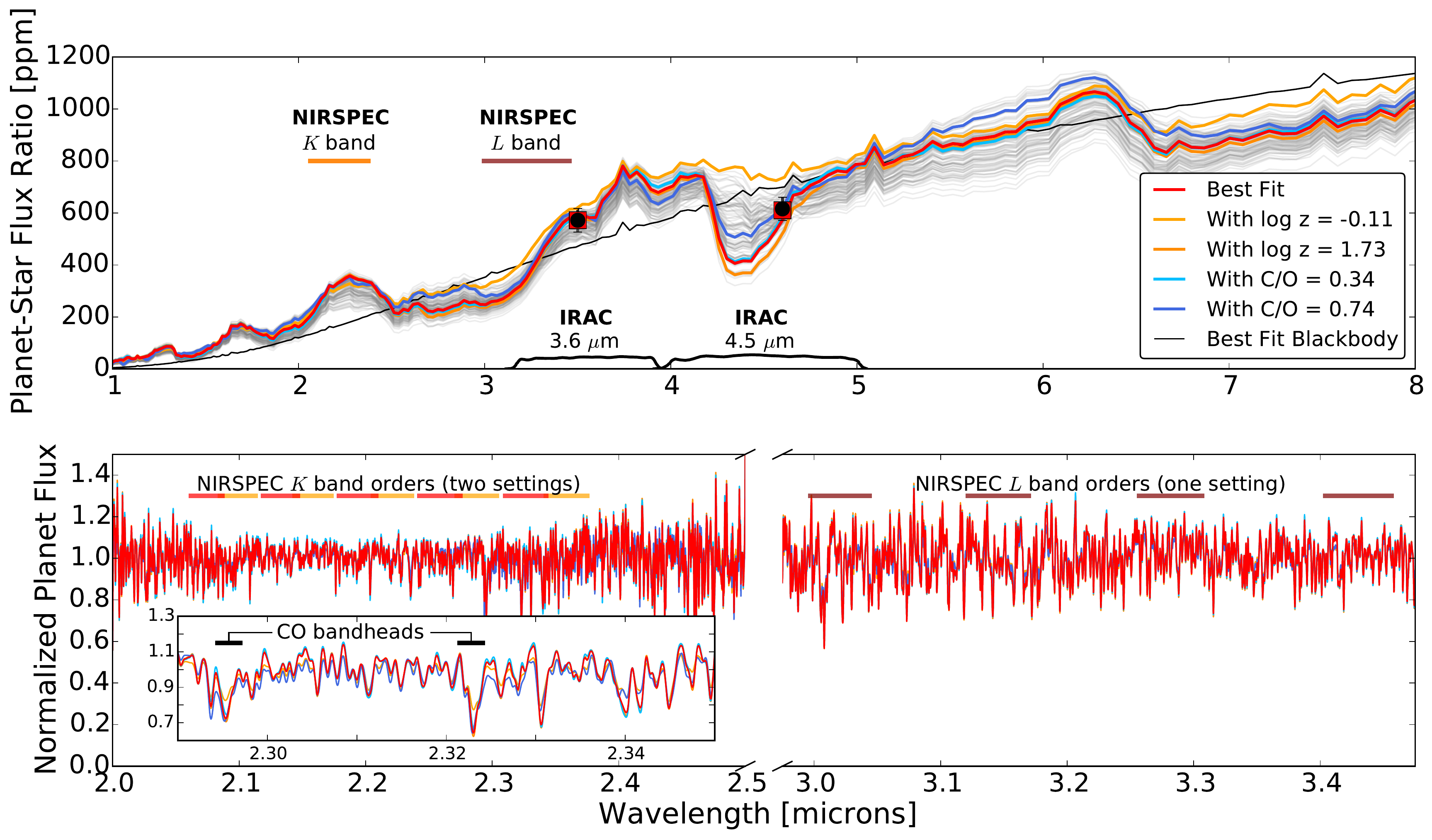} 
\caption{Model spectra plotted at probed wavelengths. Colored model spectra represent the best-fit spectrum based on the joint analysis, and spectra generated at the 16 and 84\% confidence intervals for metallicity and C/O. In the top panel, grey models represent random draws from the MCMC posterior. The best fit blackbody to the Spitzer data (1511 K, $\chi^2=8.4$) is shown in black. The Spitzer measurements and the Spitzer IRAC bandpasses are shown in black and the band-integrated best fit model fluxes are shown as red squares. The wavelength ranges of the NIRSPEC $L$ and $K$ bands are shown as well. The lower panel shows model planet spectra at NIRSPEC instrument resolution in the $L$ and $K$ bands. The $L$ band wavelength ranges observed, $K$ band wavelength ranges observed in \cite{Piskorz2017}, and CO bandheads are indicated with horizontal bars. The inset shows a wavelength region where absorption is due to CO. All other absorption is due to H$_{2}$O.}
\label{test}
\end{figure*}

The shape of the NIRSPEC-only likelihood surface (Figure~\ref{nirspecprior}) largely matches that of the uniform-prior MCMC fit to the Spitzer data (Figure~\ref{cornerplot}). Both data sets examined with uniform priors show preference for high metallicity and medium C/O ratios. The main difference in information content provided by the NIRSPEC and Spitzer data sets is at low values of redistribution. Figure~\ref{test} shows the Spitzer measurements and best-fit low- and high-resolution models at the Spitzer and NIRSPEC wavelengths investigated.

The lack of constraint on C/O is due to the opacity sources at the observed wavelengths: the NIRSPEC L-band orders are dominated by water vapor, while the Spitzer bandpasses contain water vapor, CO, and CO$_{2}$ features. Neither data set on its own can constrain C/O and the joint analysis can provide no further constraint. However the strong disagreement with models having C/O = 1 emphasizes the detection of water vapor. 

We find that our three model parameters are uncorrelated, except for $\log z$ and $f$ at high metallicity. This might result from the fact that as metallicity increases the CO and CO$_2$ abundances also increase, creating more absorption at 4.5 $\mu$m. Then, to accommodate the larger eclipse depth, the atmospheric temperature has to increase, which is essentially an increase in redistribution. 

This analysis is only for clear atmospheric models produced by the ScCHIMERA framework. We do not attempt to study clouds or hazes; this would likely require high precision data from the near-UV to mid-IR to constrain such contributions well. Thus, we are not looking at variations in the temperature profile versus molecular abundances.  The lapse rate (and therefore atmospheric composition, once C/O is set) is determined by the clear atmosphere assumption and stellar spectrum/orbital distance, and not varied.

%


When Spitzer photometry is already in hand, NIRSPEC $L$ band observations provide little new information in the $\log z$, C/O, and $f$ regions tested. While Spitzer data have more power in constraining relative abundances of species, NIRSPEC and other high resolution instruments have the ability to detect specific molecules that Spitzer lacks because of its broadband approach. Even higher resolution instruments, including CRIRES, can measure the rotation of hot Jupiters. While the spectral resolution of NIRSPEC is currently insufficient to measure hot Jupiter rotation, the situation will be markedly improved by the NIRSPEC2.0 upgrades expected to be on-sky by late CY2018. Here we find that Spitzer and NIRSPEC measurements can provide the same information independently. For example, the same SB2 analysis has been applied to NIRSPEC data of non-transiting hot Jupiters and has detected them at comparable levels of significance (e.g. 3.7$\sigma$ for upsilon Andromedae b, see \citealt{Piskorz2017}). This implies that multi-epoch NIRSPEC data of non-transiting planets is roughly equivalent to Spitzer data. With a grid of self-consistent atmospheric models and multi-epoch NIRSPEC data sets on non-transiting hot Jupiters, we may begin to constrain their line-of-sight orbital velocities, masses, inclinations and atmospheres more rigorously than in \cite{Piskorz2016, Piskorz2017}. 

There are two scenarios where NIRSPEC data may be able to provide more constraints on a hot Jupiter's atmosphere, both independently of and in tandem with Spitzer data. First, a NIRSPEC data set produced at higher signal-to-noise may provide better statistical measurements of the atmosphere's constituents. Second, a NIRSPEC data set in the $K$ band would encompass emission by carbon monoxide in the planet's atmosphere, potentially tightly pinning down the carbon-to-oxygen ratio. (See the high-resolution $K$ band models shown in the bottom panel of Figure~\ref{test}.) \citet{deKok2014} discussed how $K$, and even $H$, band high resolution spectra, in addition to $L$ band, could help constrain abundance measurements. Specifically, some of the strongest CO features are near 4.6 - 5 $\mu$m as well as 2.3 - 2.5 $\mu$m (fundamental and overtone, respectively), and so observations at the $K$ and/or $M$ bands are needed to get better C/O constraints. The best observing wavelength would depend on the temperature of the planetary atmosphere. For hot Jupiters, $K$ band data should suffice, but for more distant planets observations of the fundamental in $M$ band will likely be better. We look forward to future studies combining ground- and space-based data to probe the atmospheres of hot Jupiters and eventually Earth-size planets in the habitable zone as well. While high dispersion observations that rely on rapidly varying exoplanet radial velocities will have an increasingly more difficult time detecting distant planets (which will not move as much in a single night), this multi-epoch technique in combination with high contrast imaging will retain the ability to detect further separated planets. Other techniques are being developed to take advantage of the spatial separation of widely-separated planets \citep[e.g.][]{Snellen2014,Schwarz2016}, but these are not sensitive to planets within $\sim$0.1". Thus, the multi-epoch technique has the potential to access habitable zone planets that out of the realm of other high resolution spectroscopic techniques and as such, improvements on these types of multi-epoch techniques will be an important component of the development of high-dispersion coronography that will push to detecting biosignatures in the atmospheres of habitable zone planets. 

\section{Conclusion}
\label{conclude}
We report the ground-based detection the thermal emission of the transiting exoplanet KELT-2Ab's by measuring the planet's Doppler shift at various orbital phases. We measure $K_P$ = 148 $\pm$ \Kperror\ km/s and report the presence of water vapor in its atmosphere. The agreement of this measurement with transit and radial velocity measurements reinforces the conclusions from previous detections (e.g., \cite{Brogi2012, Brogi2013, Brogi2014, Lockwood2014, Piskorz2016, Piskorz2017}, etc.) regarding the utility of the cross-correlation technique. Rigorous exploration of the phase space near $K_P$ = 148 km/s with a suite of planetary atmospheric spectra determines the atmospheric properties required to satisfy both the Spitzer and NIRSPEC data. In the future, we will observe this planet in the $K$ and $M$ bands at high resolution.  The combination of ground-based data with space-based data will hopefully provide new constraints on the C/O ratio of the planet's atmosphere and provide additional insight into the planet's formation history.

\acknowledgments{The authors wish to recognize and acknowledge the very significant cultural role and reverence that the summit of Mauna Kea has always had within the indigenous Hawaiian community.  We are most fortunate to have the opportunity to conduct observations from this mountain. The data presented herein were obtained at the W.M. Keck Observatory, which is operated as a scientific partnership among the California Institute of Technology, the University of California and the National Aeronautics and Space Administration. The Observatory was made possible by the generous financial support of the W.M. Keck Foundation.  This work is also based on observations made with the Spitzer Space Telescope, which is operated by the Jet Propulsion Laboratory, California Institute of Technology, under a contract with NASA. This work was partially supported by funding from the NSF Astronomy \& Astrophysics and NASA Exoplanets Research Programs (grants AST-1109857 and NNX16AI14G, G.A. Blake P.I.). M.R.L would like to thank Richard Freedman and Roxana Lupu for providing pre-tabulated line-by-line absorption cross-sections, Paul Molliere, Ryan Garland, Joanna Barstow, Ingo Waldman, and Marco Rochetto for useful discussions regarding correlated-K, and Ty Robinson and Mark Marley for useful radiative-convective modeling discussions.  H.A.K. acknowledges support from the Sloan Foundation. Lastly, we thank an anonymous reviewer for insightful comments which improved the content of this paper. }

\end{document}